\documentclass[aps,showpacs,twocolumn,prl]{revtex4}
\usepackage{epsfig}
\usepackage{color}

\begin{document}

\newcommand{\pderiv}[2]{\frac{\partial #1}{\partial #2}}
\newcommand{\deriv}[2]{\frac{d #1}{d #2}}
\newcommand{\eq}[1]{Eq.~(\ref{#1})}
\newcommand{\infint}{\int \limits_{-\infty}^{\infty}}

\title{Thermostatistics of overdamped motion of interacting particles}

\author{J.~S. Andrade Jr.$^{1,3}$, G.~F.~T. da Silva$^{1}$, A.~A. Moreira$^{1}$,
F.~D. Nobre$^{2,3}$, E.~M.~F. Curado$^{2,3}$}
\affiliation{$^{1}$Departamento de F\'{\i}sica, Universidade Federal
do Cear\'a, 60451-970 Fortaleza, Cear\'a, Brazil\\ $^{2}$Centro
Brasileiro de Pesquisas F\'{\i}sicas, Rua Xavier Sigaud 150, 22290-180, 
Rio de Janeiro-RJ, Brazil\\$^{3}$National Institute of Science and Technology 
for Complex Systems, Rua Xavier Sigaud 150, 22290-180, Rio de Janeiro-RJ, Brazil}

\date{\today}

\begin{abstract}
  We show through a nonlinear Fokker-Planck formalism, and confirm by
  molecular dynamics simulations, that the overdamped motion of
  interacting particles at $T=0$, where $T$ is the temperature of a
  thermal bath connected to the system, can be directly associated
  with Tsallis thermostatistics. For sufficiently high values of $T$,
  the distribution of particles becomes Gaussian, so that the
  classical Boltzmann-Gibbs behavior is recovered. For intermediate
  temperatures of the thermal bath, the system displays a mixed
  behaviour that follows a novel type of thermostatistics, where the
  entropy is given by a linear combination of Tsallis and
  Boltzmann-Gibbs entropies.
\end{abstract}


\maketitle

Nonlinear Fokker-Planck equations (FPE's) \cite{frankbook05} are
frequently employed to represent macroscopically physical and chemical
systems displaying anomalous diffusion behavior \cite{bouchaud90}.
Scientifically and technologically important examples of such systems
include, among others, the flow through porous media \cite{muskat37},
the dynamics of surface growth \cite{spohn93}, the diffusion of
polymer-like breakable micelles \cite{bouchaud91}, and the dynamics of
interacting vortexes in disordered superconductors
\cite{zapperi01,moreira02,barrozo09}. A typical nonlinear FPE may be
written in the general form \cite{frankbook05,schwaemmle07},
\begin{equation}
\label{eq:gennlfpe}
\frac{\partial P}{\partial t}= -\frac{\partial [A(x) \Psi (P)]}{\partial x}
+\frac{\partial}{\partial x} \left[ \Omega (P) \frac{\partial P}{\partial x} \right]~,
\end{equation}
where the external force $A(x)$ is associated with a potential
$\phi(x)$ [$A(x)=-d\phi(x)/dx$], and the analyticity of the potential
as well as the integrability of the force are assumed to hold in all
space. The functionals $\Psi [P(x,t)]$ and $\Omega [P(x,t)]$ satisfy
requirements of positiveness, integrability, and differentiability
with respect to $P(x,t)$ \cite{schwaemmle07}. Moreover, in order to
preserve the probability normalization for all times, one should
impose the probability distribution, together with its first
derivative, as well as the product $A(x)\Psi[P(x,t)]$, to be zero in
the limits $x \rightarrow \pm \infty$.

An important result associated with nonlinear FPE's is the H-theorem
and its generalizations
\cite{frankbook05,shiino01,frankdaff01,chavanis03,schwaemmle07,schwaemmle09}.
In the case of a system subjected to an external potential, the
H-theorem leads to a well-defined sign for the time derivative of the
free-energy functional,
\begin{equation}
\label{eq:free_energy}
F = U - \gamma S~;
\qquad U =  \int_{-\infty}^{\infty}dx \ \phi(x) P(x,t)~,
\end{equation}
where $\gamma$ represents a positive Lagrange multiplier and the
entropy may be considered in a very general form as,
\begin{equation}
\label{eq:entropysp}
S[P]\!=\!\int_{-\infty}^{\infty}\!\!g[P(x,t)]dx;~
g(0)\!=\!g(1)\!=\!0;~\frac{d^{2}g}{dP^{2}}\!\leq\! 0~,
\end{equation}
with the condition that $g[P(x,t)]$ should be at least twice
differentiable. Considering the FPE~(\ref{eq:gennlfpe}), for $dF/dt
\leq 0$ \cite{schwaemmle07,schwaemmle09}, we obtain,
\begin{equation}
\label{eq:genrelation}
-\gamma \deriv{^2 g[P]}{P^2}=\frac{\Omega[P]}{\Psi[P]}~.
\end{equation}
A relevant outcome of \eq{eq:genrelation} is that the ratio
$\Omega[P]/\Psi[P]$ determines an entire class of FPE's associated
with a single entropic form \cite{schwaemmle07,schwaemmle09}. Here we
are interested on the following type of nonlinear FPE
\cite{plastino95,tsallis96}:
\begin{equation}
\label{eq:tsallisnlfpe}
\frac{\partial P}{\partial t}= -\frac{\partial[A(x)P]}{\partial x}
+D\nu \frac{\partial}{\partial x} 
\left[ P^{\nu-1} \frac{\partial P}{\partial x} \right]~,
\end{equation}
where $D$ is a constant, $\nu$ is a real number and the functionals in
\eq{eq:gennlfpe} correspond to $\Psi[P(x,t)]=P(x,t)$ and
$\Omega[P(x,t)]=D\nu[P(x,t)]^{\nu-1}$. By substituting these quantities in
\eq{eq:genrelation}, integrating and using the conditions
(\ref{eq:entropysp}), one obtains Tsallis entropy
\cite{tsallis88,tsallis09,schwaemmle07}, for which,
\begin{equation}
\label{eq:tsallisgp}
g[P] = k \ {[P(x,t)]^{\nu} - P(x,t) \over 1-\nu}~,
\end{equation}
where $k\equiv D/\gamma$. Considering an external force
$A(x)\!=\!-\alpha x$ ($\alpha \ge 0$), the solution for
\eq{eq:tsallisnlfpe} with initial condition $P(x,0)\!=\!\delta(x)$ is
given by the distribution,
\begin{equation}
\label{eq:tsallisdist}
P(x,t)=B(t) [ 1 + \beta (t)(1-\nu)x^{2} ]_{+}^{1/(\nu-1)}~, 
\end{equation}
where $[y]_{+}\!=\!y$, for $y>0$ and zero otherwise. The
time-dependent parameters $B(t)$ and $\beta (t)$ are defined in such a
way as to preserve the norm and the form of the distribution for all
times \cite{plastino95,tsallis96,curado03}.
Equation~(\ref{eq:tsallisdist}) corresponds exactly to the
distribution obtained through extremization of Tsallis entropy
\eq{eq:tsallisgp}, for the energy defined as in
Eq.~(\ref{eq:free_energy}), and under standard constraints of
probability normalization. It is important to notice that, by
substituting $\nu=2-q$ in \eq{eq:tsallisdist}, one recovers precisely
the usual distribution of nonextensive statistical mechanics, known as
$q$-Gaussian, as obtained through a more general definition for the
internal energy \cite{tsallis09}.
       
Additionally, the way in which the system evolves dynamically, as well
as its stationary state, provide distinctive signatures of anomalous
diffusion behavior. For small times $t \ll 1$, it is possible to show
that the diffusion propagation front $x(t)$ dictated by
\eq{eq:tsallisnlfpe} advances as $\langle x(t) \rangle \propto
t^{2/(\nu+1)}$ \cite{schwaemmle09}. In the limit $t \rightarrow
\infty$, the system approaches the stationary state given by,
\begin{equation}
\label{eq:statparameters}
P_{\text{st}}(x)=B^{*}[1-\beta^{*} x^{2}]_{+}^{1/(\nu-1)}~,
\end{equation}
depending on the initial time $t_{0}$,
$\beta^{*}\!=\!\beta(t_{0})[B^{*}/B(t_{0})]^{2}$, and
$B^{*}\!=\![\alpha B(t_{0})^{2}/2D\nu \beta(t_{0})]^{1/(1+\nu)}$.  
One should stress that this form of stationary distribution holds 
for any confining potential $\phi(x)$, by simply replacing in
Eq.~(\ref{eq:statparameters}) the term $x^2$ with $\phi(x)$
\cite{plastino95}.

Next we show that the microscopic behavior of a system of interacting
particles undergoing overdamped motion is fully compatible with a
continuum nonlinear diffusion equation. Moreover, we find that such a
continuum formulation for a highly dissipative system corresponds to
the FPE (\ref{eq:tsallisnlfpe}) with $\nu=2$. We start by considering
the equation of motion for a particle $i$ in a system of $N$
overdamped particles,
\begin{equation}
\mu \vec{v}_{i}=\sum_{j \neq i}\vec{J}(\vec{r}_{i}-\vec{r}_{j})+
\vec{F}^{e}(\vec{r}_{i})+\eta(\vec{r}_{i},t)~,
\label{eq:mov}
\end{equation}
where $\vec{v}_{i}$ is the velocity of the $i$th particle, $\mu$ is
the effective viscosity of the medium, the first term on the right
accounts for the interactions among particles,
$\vec{F}^{e}(\vec{r}_{i})$ represents an external force, and $\eta$
corresponds to an uncorrelated thermal noise with zero mean and
variance $\langle \eta^2\rangle\!=\!k_{B}T/\mu$. Here we consider a
short-range repulsive particle-particle interaction in the form,
$\vec{J}(\vec{r})\equiv {G}(|\vec{r}|/\lambda)\hat{r}$, where
$\hat{r}$ is the unit vector along the axis connecting each pair of
particles, and $\lambda$ is a characteristic length of the short-range
pairwise interaction.

To obtain a continuum description of this system
\cite{zapperi01}, we perform a coarse graining of Eq.~(\ref{eq:mov}),
starting from the Fokker-Planck equation for the probability
distribution of the particle coordinates 
${\cal P}(\vec{r}_1,....,\vec{r}_N,t)$,
\begin{equation}
\mu\frac{\partial {\cal P}}{\partial t}=\sum_{i}\vec{\nabla}_{i}(-\vec{f}_{i}
{\cal P}+k_{B} T\vec{\nabla}_{i}{\cal P})~,
\label{eq:fp}
\end{equation}
where $\vec{f}_i$ is the force on the particle $i$ given by 
Eq.~(\ref{eq:mov}). By introducing the single particle density
$\rho(\vec{r},t)\equiv\langle\sum_{i}\delta^{2}(\vec{r}-\vec{r}_i)\rangle$, 
where the average is made over the distribution 
${\cal P}(\vec{r}_1,....,\vec{r}_N,t)$, one obtains, 
\begin{eqnarray}
\mu\frac{\partial \rho}{\partial t} & = &
-\vec{\nabla}\big[\int d^2r' \vec{J}(\vec{r}-\vec{r}\,')
\rho^{(2)}(\vec{r},\vec{r}\,',t) \nonumber \\ &  &
+\vec{F}^{e}(\vec{r})\rho(\vec{r},t)\big]
+k_{B}T \nabla^{2}\rho(\vec{r},t)~,
\end{eqnarray}
where $\rho^{(2)}(\vec{r},\vec{r}\,',t)$ is the two-point density.
If we now assume that the approximation $\rho^{(2)}(\vec{r},\vec{r}\,',t)
\simeq \rho(\vec{r},t)\rho(\vec{r}\,',t)$ is valid, we can then 
coarse grain the particle-particle interaction force to obtain
\cite{zapperi01},
\begin{equation}
\int\! d^2r' \vec{J}(\vec{r}-\vec{r}\,')\rho(\vec{r}\,',t)\!\simeq\! 
-a\vec{\nabla} \rho(\vec{r},t)~;~a\!\equiv\! \int\!d^2r
\vec{r}\cdot\vec{J}(\vec{r})/2~,
\label{eq:a}
\end{equation}
where only length scales larger than $\lambda$ were considered.

Here we investigate the motion of particles in a two-dimensional
narrow channel of size $L_{x} \times L_{y}$ under an external force in
the $x$-direction, $\vec{F}^{e}=-A(x)\hat{x}$.  We then assume that
the concentration is only weakly dependent on the transverse
$y$-coordinate, $\rho(\mathbf{r},t) \approx \rho(x,t)$.  After
collecting all force terms, we obtain,
\begin{equation}
\label{eq:rhodiffusioneqT}
\mu \frac{\partial \rho}{\partial t} = \frac{\partial}{\partial x} 
\left\{ \rho \left[ a \ \frac{\partial \rho}{\partial x}
- A(x) \right] \right\} + k_{B}T \frac{\partial^{2} \rho}{\partial x^{2}}~.
\end{equation}
Interestingly, by introducing the probability
$P(x,t)\!=(L_{y}/N)\rho(x,t)$, defining $D\!=\!(aN)/(\nu L_{y})$, and
setting $\mu=1$ and $T=0$, we recover precisely the
FPE~(\ref{eq:tsallisnlfpe}) for $\nu=2$. Herein we restrict our study
to a restoring harmonic force, $A(x)=-\alpha x$ ($\alpha \ge 0$). In
the limit $T=0$, the steady-state solution of
Eq.~(\ref{eq:rhodiffusioneqT}) is given by
\begin{equation}
\label{eq:q2statparameters}
\rho_{st}=\frac{\alpha}{2a}(x_e^2-x^2);~|x|<x_e~,
\end{equation}
with $x_e=(3Na/2\alpha L_y)^{1/3}$.  This result is identical to
\eq{eq:statparameters} if there one adopts $\nu=2$ and
$\beta^{*}\!=\!x_{e}^{-2}$.

We now show through direct molecular dynamics (MD) simulations that
the behavior of a typical overdamped system at $T=0$ is fully
compatible with the solution (\ref{eq:q2statparameters}), therefore
representing a microdynamical realization of Tsallis thermostatistics.
As an example, we consider a system of vortex lines moving on a type
II superconductor substrate~\cite{nori91,bryskin93}, but the results
presented here could nevertheless be extended to other forms of
short-range repulsive interactions~\cite{barrozo09}. In this case, the
effective viscosity is given by $\mu=\Phi_0 H_{c2}/\omega c^2$, where
$\Phi_0$ is the magnetic quantum flux, $c$ is the speed of light,
$\omega$ is the resistivity of the normal phase, and $H_{c2}$ is the
upper critical field. Moreover, the commonly adopted vortex-vortex
interaction is $\vec{J}(\vec{r})\equiv
f_{0}K_{1}(|\vec{r}|/\lambda)\hat{r}$, where $K_{1}$ is a modified
Bessel function decaying exponentially for $|\vec{r}| > \lambda$, the
pre-factor is given by $f_{0}=\Phi_{0}^2/(8\pi\lambda^3)$, and
$\lambda$ corresponds to the London penetration
length~\cite{degennes66}.

\begin{figure}[t]
\begin{center}
\includegraphics*[width=8cm]{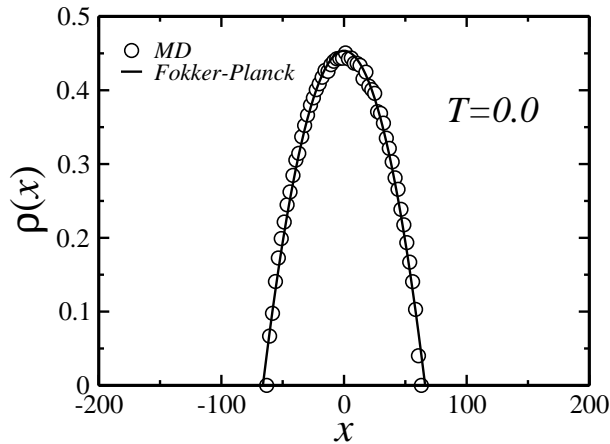}
\end{center}
\caption{Profile of the density of particles at stationary state
and $T=0$, obtained from molecular dynamics by integrating \eq{eq:mov}
(empty circles), as compared to the theoretical estimate
\eq{eq:q2statparameters}. The best fit to the data of a quadratic
function gives the parameter $a=2.41 f_{0} \lambda^{3}$ (full
line). The position $x$ is measured in units of $\lambda$, whereas the
steady-state density $\rho(x)$ is expressed in units $\lambda^{-2}$.}
\label{fig:partdensity}
\end{figure}

We perform MD simulations with $N=800$ flux lines placed at random
within a two-dimensional channel of sizes $L_{x}\!=\!100 \lambda$,
$L_{y}\!=\!20 \lambda$, where periodic boundary conditions are imposed
in the $y$-direction. In these simulations, we first set $T\!=\!0$ and
a confining external force is applied in the $x$-direction with
restoring constant $\alpha = 10^{-3} f_{0}/\lambda$. The equations of
motion (\ref{eq:mov}) are numerically integrated and the system
evolves in time until a stationary state of mechanical equilibrium is
reached, which is identified here in terms of an invariant density profile.
As shown in Fig.~\ref{fig:partdensity}, the obtained stationary
profile of particle density is clearly parabolic in shape, in perfect
agreement with the theoretical prediction of \eq{eq:q2statparameters}.
The least-squares fit to the simulation data of a quadratic function
results in the parameter $a\approx 2.41 f_{0}\lambda^{3}$. Strictly
speaking, the predicted value $a=\pi~f_{0}\lambda^{3}$, calculated
from \eq{eq:a}, is valid only for unconfined systems, if we assume
that the density $\rho$ varies slowly within the interaction range of
a particle. The discrete character of the interacting particles
therefore leads to a correction on this theoretical prediction. In the
absence of external forces, i.e., for $\alpha=0$, our results show
that the front propagation at early times evolves as $\langle x(t)
\rangle \propto t^{1/3}$ \cite{zapperi01}. Once more, the
compatibility between this dynamical scaling and the anomalous
diffusion behavior intrinsically associated with \eq{eq:tsallisnlfpe}
for $\nu=2$ confirms the validity of our approach.

Having shown that overdamped particles in the limit of $T=0$, typified
as interacting flux lines on a type II superconductor substrate, obeys
Tsallis statistics with an entropic index $\nu=2$, we now analyze the
effect of finite $T$ on this system. From \eq{eq:rhodiffusioneqT}, one
can envisage the competition between two types of diffusion, which are
associated, respectively, with the strength of interactions between
vortexes, controlled by the parameter, $a$, or equivalently, $D$, and
the temperature of the thermal bath, $T$. In this way, the ratio
$k_{B}T/a$ plays a crucial role in the time evolution of the system.
As for the case $T=0$, in the absence of external forces, i.e., for
$\alpha=0$, the diffusion behavior of the system for $k_{B}T\! \ll \!
a$ should be governed by the anomalous features associated with the
index $\nu=2$, namely, $\langle x^{2} \rangle \propto t^{2/3}$,
whereas normal diffusion prevails for $k_{B}T\! \gg \! a$, i.e.,
$\langle x^{2} \rangle \propto t$. In the presence of a restoring
external force and for $T>0$, a stationary-state analytical solution
for \eq{eq:rhodiffusioneqT} can still be obtained,
\begin{equation}
\label{eq:solutionlambert}
\rho(x) = \frac{k_{B}T}{a} \ W \left\{ \frac{a \rho(0)}{k_{B}T} \
\exp \left[ \frac{a \rho(0)}{k_{B}T} -
\frac{\alpha x^{2}}{2k_{B}T} \right]
 \right\}~,
\end{equation}
where the $W$-Lambert function is defined implicitly through the
equation $W(z)e^{W(z)}=z$ (see \cite{valluri2009} and references therein).
\begin{figure}[t]
\begin{center}
\includegraphics*[width=8cm]{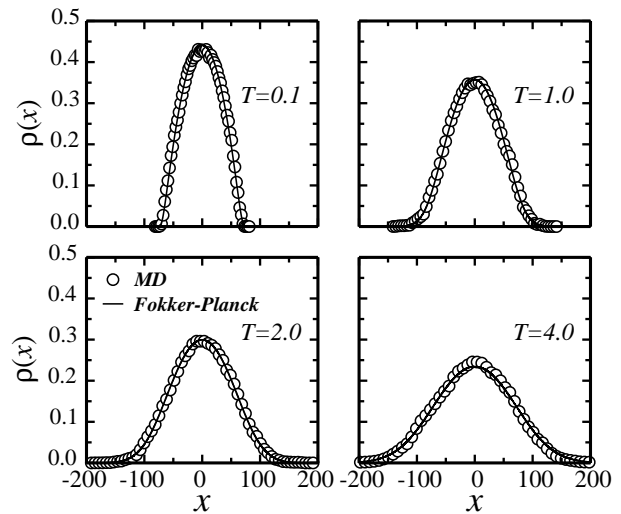}
\end{center}
\protect\caption{Comparison between the density of particles profiles
  in the stationary state obtained from molecular dynamics (empty
  circles) and the theoretical prediction \eq{eq:solutionlambert}
  (full lines), for different values of the temperature of the thermal
  bath, $T$. The position $x$ is measured in units of $\lambda$, the
  density is in units $\lambda^{-2}$, and $T$ is given in units
  $(f_0\lambda/k_{B})$. The parameter $a\!=\!2.41 f_{0}\lambda^{3}$ is
  the same adopted for $T\!=\!0$, whereas $\rho(0)$ is determined by
  conservation of the total number of particles in the system,
  $\int_{-L_{x}/2}^{L_{x}/2}\rho(x)dx=(N/L_{y})$. For low values of $T$,
  e.g., $T=0.1f_{0}\lambda/k_B$, the density profile is approximately
  parabolic. At intermediate values of $T$, the system can be
  described by a $W$-Lambert function \eq{eq:solutionlambert}. At high
  values of $T$, the profile becomes typically Gaussian, as illustrated
  from results at $T\!=\!4f_{0}\lambda/k_B$.}
\label{fig:partdensitytemp}
\end{figure}
In order to test this prediction, extensive MD simulations have been
performed for different values of $T$. As depicted in
Fig.~\ref{fig:partdensitytemp}, we find very good agreement between
the continuum model solution \eq{eq:solutionlambert}, and MD results
of density profiles at steady-state. For consistency, we consider
the same numerical value used to adjust the profile at $T=0$,
namely, $a=2.41 f_{0}\lambda^{3}$, while the parameter $\rho(0)$ in
\eq{eq:solutionlambert} has been determined by a conservation constraint 
in the total number of particles, $\int_{-L_{x}/2}^{L_{x}/2}\rho(x)dx=(N/L_{y})$.  

The effect of increasing $T$ is to gradually change the density
profile from a parabolic to a Gaussian shape. Indeed these two limits
correspond, respectively, to the particular cases of Tsallis
distribution with index $\nu=2$ and the standard Boltzmann-Gibbs
statistics. Next we show that the profiles obtained from
\eq{eq:solutionlambert}, for finite $T$, are in fact associated with a
novel type of entropy form. By comparing Eqs.~(\ref{eq:rhodiffusioneqT}) 
and (\ref{eq:gennlfpe}), we obtain that $\Psi[P(x,t)]\!=\!P(x,t)$ and
$\Omega[P(x,t)]\!=\!2DP(x,t)+k_{B}T$.  Substituting these quantities
into \eq{eq:genrelation}, integrating twice and using the conditions
(\ref{eq:entropysp}), one gets,
\begin{eqnarray}
\label{eq:gpgeneral}
g[P(x,t)] & = & \frac{D}{\bar{\gamma}}[P(x,t)-P^{2}(x,t)] \nonumber \\ &  &
- \frac{k_{B}T}{\bar{\gamma}}[P(x,t) \ln P(x,t)]~, 
\end{eqnarray}
where $\bar{\gamma}$ is a positive Lagrange parameter defined through
a free-energy functional like the one given by \eq{eq:free_energy},
and $x$ is a conveniently rescaled variable. This functional leads to
the following entropic form:
\begin{eqnarray}
\label{eq:spgeneral}
S[P] & = & \frac{D}{\bar{\gamma}} [1 - \int_{-\infty}^{\infty}dx 
\ P^{2}(x,t)] \nonumber \\ &  &
- \frac{k_{B}T}{\bar{\gamma}} 
\int_{-\infty}^{\infty}dx \ P(x,t) \ln P(x,t)~.   
\end{eqnarray}
Equation~(\ref{eq:spgeneral}) is precisely the sum of Tsallis entropy
with $\nu=2$, which appears as a consequence of many-body
interactions, and Boltzmann-Gibbs entropy, which comes from the
thermal noise of the bath. The entropy is zero only for $D=T=0$. One
may now extremize this mixed entropy under the constraints,
$\int_{-\infty}^{\infty}P(x,t)dx=1$ and
$U=\int_{-\infty}^{\infty}[\phi(x)-\phi_{0}]P(x,t)dx$, by defining the
functional,
\begin{eqnarray}
\label{eq:sigmafunc}
\Sigma[P(x,t)] & = & \frac{\bar{\gamma}}{k_{B}T} \ S[P]
- a_{1} \int_{-\infty}^{\infty}dx \ P(x,t) \nonumber \\ &  &
- a_{2} \int_{-\infty}^{\infty}dx \ [\phi(x)-\phi_{0}]P(x,t)~. 
\end{eqnarray}
The extreme condition $\delta\Sigma/\delta P=0$ finally leads to
\begin{eqnarray}
\label{eq:relpx}
&\!\!\!\!\!\!\!\!\!\!\!\!\!\!\!\!\!\!\!\!\!\!\!\!\!\!\!\!\!\!\!\!\!\!\!\!
P(x) \exp \left[ \frac{2D}{k_{B}T}  \ P(x)
\right] = \nonumber \\ 
& \;\;\;\;\;\;\exp \left\{-(1+a_{1})-a_{2}[\phi(x)-\phi_{0}] \right\}~,   
\end{eqnarray}
which may be written in terms of $W$-Lambert's function, i.e., in the 
same form as \eq{eq:solutionlambert}, by considering a harmonic potential
$\phi(x)=\alpha x^{2}/2$, multiplying both sides by $2D/(k_{B}T)$,
identifying $a_{2}=1/(k_{B}T)$, defining $z$ as the right term in
\eq{eq:relpx}, and $W(z)=2DP(x)/(k_{B}T)$. This confirms that the
entropy \eq{eq:spgeneral} is directly related to the stationary
solution \eq{eq:solutionlambert} of the FPE
(\ref{eq:rhodiffusioneqT}).
       
In summary, we have shown through a nonlinear Fokker-Planck formalism
and confirmed through MD simulations that a system of interacting
particles undergoing overdamped motion at $T=0$, where $T$ is the
temperature of a thermal bath connected to the system, can be
considered as a physical realization of Tsallis thermostatistics with
an entropic index $\nu=2$ in \eq{eq:tsallisgp}. At high values of $T$,
the classical Boltzmann-Gibbs behavior is recovered. At intermediate
values of $T$, our approach, also confirmed by MD simulations, leads
to a stationary solution for the corresponding nonlinear FPE that can
be expressed in terms of the $W$-Lambert function. As a consequence,
we disclose a novel mixed entropic form based on a linear combination
between Tsallis and Boltzmann-Gibbs entropies.

We thank the Brazilian agencies CNPq, CAPES, FAPERJ and FUNCAP, and
the CNPq/FUNCAP-Pronex grant for financial support. We also thank
Constantino Tsallis for fruitful discussions.

\end{document}